\documentclass[prl,aps,twocolumn,superscriptaddress,floatfix,nobibnotes,amsbsy]{revtex4-2}

\usepackage{epsfig,grffile}
\usepackage{graphicx,graphics}
\usepackage[table]{xcolor}
\usepackage{epstopdf}
\usepackage{amsmath,amssymb,amsfonts,stmaryrd,amsthm,mathrsfs}
\usepackage{latexsym,verbatim}
\usepackage{bm}
\usepackage{bbold}
\usepackage{color}
\usepackage{ulem}
\usepackage[breaklinks=false,colorlinks,citecolor=blue,linkcolor=blue,urlcolor=blue]{hyperref}
\usepackage[capitalise]{cleveref}
\usepackage[shortlabels]{enumitem}
\usepackage[abs]{overpic}
\usepackage{textcomp} 
\usepackage{mathtools}
\usepackage{braket}
\usepackage{comment}
\usepackage{tcolorbox}
\usepackage{dutchcal}
\usepackage{soul}    
\usepackage{orcidlink}

\DeclareMathOperator{\tr}{tr}

\DeclareMathOperator{\diag}{diag}

\usepackage{physics}
\usepackage{tensor}
\usepackage{abbrevs}
\usepackage[capitalise]{cleveref}

\newname\bh{$\bar{H}$}
\newname\verified{{\color{green} \maltese }}
\newname\hamt{$ \frac{i}{\hbar} \ham _0 t $}
\newcommand{\ham}{\hat H}

\DeclareFontFamily{U}{mathx}{\hyphenchar\font45}
\DeclareFontShape{U}{mathx}{m}{n}{
      <5> <6> <7> <8> <9> <10>
      <10.95> <12> <14.4> <17.28> <20.74> <24.88>
      mathx10
      }{}
\DeclareSymbolFont{mathx}{U}{mathx}{m}{n}
\DeclareFontSubstitution{U}{mathx}{m}{n}
\DeclareMathAccent{\widecheck}{0}{mathx}{"71}
\DeclareMathAccent{\wideparen}{0}{mathx}{"75}

\begin{document}
 
\title{ Quantum dynamics in the self-consistent quadratic approximation}

\author{Frank Ernesto Quintela Rodriguez\,\orcidlink{0000-0002-9475-2267}}
\email[Email: ]{frank.quintelarodriguez@sns.it}
\affiliation{Scuola Normale Superiore, Piazza dei Cavalieri, 7, Pisa, 56126, Italy}

\begin{abstract} 
A self-consistent quadratic theory is presented to account for nonlinear contributions in quantum dynamics. 
Evolution equations are shown to depend on higher-order gradients of the Hamiltonian, which are incorporated via their equations of motion or through perturbative calculations.
The dynamics is proven trace-preserving, with the Hamiltonian acting as a constant of motion for initial Gaussian states.
Nonlinear response functions are calculated perturbatively, and sufficient conditions are provided for the existence of their classical limit.
\end{abstract}




\date{\today}

\maketitle

We are concerned with nonlinear contributions in quantum dynamics. 
In the second order, the dynamics consists of free evolutions under suitable unitary transformations~\cite{xiao2009theory}, or possess analytical solution under very general conditions~\cite{dodonov2003theory}.
Higher-order contributions can be incorporated via diagrammatic perturbation theory~\cite{stefanucci2013nonequilibrium} or through non-perturbative semiclassical approximations such as Gaussian wave-packet dynamics~\cite{heller1975time}, time-dependent self-consistent field~\cite{gerber1982time}, Ehrenfest dynamics~\cite{li2005ab}, self-consistent phonon formalism~\cite{shukla1981self}, stochastic self-consistent harmonic approximation~\cite{monacelli2021stochastic}, and time-dependent self-consistent harmonic approximation (TD-SCHA)~\cite{monacelli2021time,siciliano2023wigner}, to mention a few.
Here, we generalize the TD-SCHA to the case where the starting Hamiltonian is an arbitrary function of the coordinate and momentum operators, i.e., does not necessarily separates the momenta and coordinate into kinetic and potential terms, but allows for a mixing of them.
This can be done by exploiting the Wigner-Weyl isomorphism, which maps a Hilbert space operator ${A \in \mathscr H _d}$ into a phase space distribution ${\mathcal A(z),\, z\in \mathbb R^{2d}}$, called its Weyl symbol~\cite{gadella1995moyal,de2017wigner}.

Let's consider some representative Hamiltonians that, although  not explicit functions of the coordinate and momentum operators, can be recast as such.
It is known that any operator in the Lie algebra of $SU(m,n)$ can always be expressed in terms of boson operators~\cite{barry2008qubit,leverrier2018p,puri1994m,puri2001mathematical}.
This is further extended by tensor product operations and coupling to multi-mode boson operators ${\{ \mathbf a^\dagger,\, \mathbf a \} }$ through
\begin{align} \label{m_145}
    & H (t) =  \sum_j f_j(\mathbf Y , t) \otimes g_j ( \mathbf a^\dagger , \mathbf a , t ) ,
\end{align}
where ${\mathbf Y_k \in SU(m,n)} $ are the generators of the Lie algebra, and $f_j,\, g_j$ arbitrary functions.
In the coordinate-momentum representation $ H(t)$ becomes $ H(q,t) $, where ${ q = (p_1,\dots, p_n, x_1,\dots ,x_n) ^\mathrm T}$ is hereafter referred to as the \textit{canonical~vector}.
When ${ \mathbf Y_k \in SU(d,0) \equiv SU(d) }$, \cref{m_145} describes the interaction of a $d$-level quantum system (\textit{qudit}) with a boson field, as illustrated by the Jaynes-Cummings Hamiltonian~\cite{larson2022jaynes}. 
For these cases, an alternative formulation of the Wigner-Weyl isomorphism is presented in~\cite{tilma2016wigner,rundle2017simple,rundle2019general}.

Here, it is assumed that, for any operator $A(t)$, we are able to determine its Weyl symbol $\mathcal A_t(z)$ under the conditions outlined above.
Explicit time dependence will be omitted, such that ${A(t) \equiv A_t \equiv A}$, except when necessary for emphasis.
Once in the phase space, we utilize known results for quadratic Hamiltonians to incorporate nonlinear contributions in a self-consistent manner, hereafter called the self-consistent quadratic approximation (SCQA).  
Within this framework, equations of motions are derived for $A(t)$ and $\expval{A}_t$.
A perturbative expansion is developed to compute arbitrary correlation functions, focusing on the nonlinear response function.
This is due to its central role in coherent multidimensional spectroscopy, a cornerstone technique for studying quantum phenomena~\cite{biswas2022coherent,dorfman2016nonlinear,leone2023probing}.

Nonlinear response functions are calculated in the time-domain representation, where the SCQA yields analytical expressions provided that the initial state is Gaussian or a superposition of such states (e.g., Schrödinger cat states)~\cite{dodonov2003theory}.
A similar approach was implemented for simpler models using the coherent state representation~\cite{quintela2022vibrational,troiani2023vibrational,quintela2023vibrational}, where no truncation of the infinite-dimensional Hilbert space or the inversion of the propagator matrix was required.
Here, connections with quantum tomography are discussed for light-matter interactions of exponential form.
Furthermore, the classical limit is studied under the assumption of semiclassical admissibility of the Weyl symbols, which prevent divergences in the nonlinear response.
All of the above naturally applies to quadratic Hamiltonians with no self-consistency conditions, for which, to the author's knowledge, arbitrary-order nonlinear response functions have not been calculated elsewhere.
Quadratic Hamiltonians are often encountered in coherent spectroscopy, as they characterize the lower-order contributions of the interaction potentials.

We start with the quantum evolution in the Schrödinger picture, 
\begin{align} \label{m_14}
    & \rho _t =  U _t \rho_0  U^\dagger _t \\
    \label{m_15}
    & U _t = \mathcal T e^{ -i /\hbar \int _0 ^t  H (\tau ) \dd \tau } \\
    & \expval{ A}_t = \tr {  A \rho _t }, 
\end{align}
where the symbol $\mathcal T$ indicates the synchronously time-ordered exponential~\cite{mukamel1995principles}.
In the SCQA, the Hamiltonian $H(t)$ is approximated by the quadratic form 
\begin{align} \label{m_13}
    &  H_{ \text{SC}} (t) = \frac{1}{2}  q ^\mathrm T  B_t  q +  q ^\mathrm T  C_t ,
\end{align}
where 
\begin{align} \label{m_72}
    &  B_t = \expval{ \nabla ^2 H }_t \\
    \label{m_143}
    &  C_t = \expval{ \nabla H}_t -   \expval{ \nabla ^2 H } _t  \expval{  q} _t .
\end{align}
We denote by ``$\nabla ^\alpha H$'' the operator associated with the Weyl symbol $\nabla ^\alpha \mathcal H (z)$.
Its expectation value is given by
\begin{align} \label{m_138}
    & \expval {\nabla ^\alpha H} _t = \tr { \nabla ^\alpha H \rho_t } = \int_{\mathbb R^{2n}} \dd z\, \nabla ^\alpha \mathcal H ( z) \mathcal W_t (z) ,
\end{align}
where the Weyl symbol of $\rho_t$, denoted as $\mathcal W_t (z)$, is commonly referred to as the \textit{Wigner function}.
The derivatives $ \nabla ^\alpha $  are well-defined when $\mathcal H(z)$ is a generalized function, provided that $\mathcal W_t (z) $ is continuously differentiable.
Under the $H_{SC} (t)$ Hamiltonian, the evolution operator is approximated by 
\begin{align} \label{m_63}
    & U _t  \approx U_{SC} (t) = \mathcal T e^{ -i /\hbar \int _0 ^t  H_{\text{SC}} (\tau ) \dd \tau } .
\end{align}
The term \textit{self-consistent} results from using the evolved state $\rho_t$ in the $H_{SC} (t)$ definition through \cref{m_14,m_138}. 
Being $H_{SC}(t)$ a quadratic form, the generated time-evolution can be described in terms of the integral of motion~\cite{gadella1995moyal,dodonov2003theory}
\begin{align} \label{m_62}
    &   q _t =  U_{SC} (t)\, q\,  U_{SC} ^\dagger (t) = \Lambda _t   q + \Delta _t ,
\end{align}
with the coefficients verifying
\begin{align} \label{m_18}
    & \dot{  \Lambda} _t =  \Lambda _t J  B_t , \quad \Lambda _0 =  E  \\
    \label{m_20}
    & \dot{ \Delta} _t =  \Lambda _t  J  C_t , \quad  \Delta _0 =  0.
\end{align}
Here $  J = \begin{pmatrix}  0 &  I \\ -  I &  0 \end{pmatrix}$ is the standard symplectic matrix and $E = \begin{pmatrix} I & 0 \\ 0 & I \end{pmatrix} $ the identity in $ \mathcal M (\mathbb R^{2n}) $.
In the Heisenberg picture, the canonical vector reads~\cite{akhundova1982wigner,dodonov2003theory}
\begin{align} \label{m_146}
    & q_H (t) \equiv U_{SC} ^\dagger (t)\, q\,  U_{SC} (t)  = \Lambda^{-1} _t  ( q - \Delta _t) ,
\end{align}
with expectation value 
\begin{align} \label{m_100}
    & \expval{  q} _t =\Lambda ^{-1}_t ( \expval{   q}_0 - \Delta _t) .
\end{align}
\Cref{m_18,m_20} apply both for the classical and quantum evolutions. 
This is a consequence of the quadratic form of the Hamiltonian, for which the Poisson bracket coincides with the Moyal commutator ${ \acomm{H}{f}_P = \acomm{H}{f}_M }$~\cite{gadella1995moyal}.
The state of the system is entirely specified by $\Lambda _t,\, \Delta _t $, e.g., the initial Wigner function $ \mathcal W _0 ( z) $ evolves according to~\cite{dodonov2003theory}
\begin{align} \label{m_38}
    &  \mathcal W ( z,\, t) = \mathcal  W_0 ( \Lambda _t  z + \Delta _t) .
\end{align}
The matrix $\Lambda_t$ is symplectic, hence the identities 
\begin{align}  \label{m_73}
    & \Lambda  ^\mathrm T _t  J \Lambda _t =   J \\
    \label{m_90}
    & \Lambda ^{-1} _t =   J \Lambda ^\mathrm T _t  J ^\mathrm T. 
\end{align}
Collecting the above results, we have the SCQA equations 
\begin{subequations} \label{m_80}
\begin{align} 
    & \dot{  \Lambda} _t = \Lambda _t J  B_t , \quad \Lambda _0 =  E\\
    \label{m_81}
    & \dot{ \Delta} _t =  \Lambda _t  J  C_t, \quad \Delta _0 =  0 \\
    \label{m_83}
    &  B_t = \expval{ \nabla ^2 H }_t \\
    \label{m_84}
    &  C_t = \expval{ \nabla H}_t -   \expval{ \nabla ^2 H } _t  \expval{  q} _t \\
    \label{m_82}
    & \expval{  q} _t =  \Lambda ^{-1}_t ( \expval{   q}_0 - \Delta _t) .
\end{align}
\end{subequations}
To complete these equations, we must specify the quantities $\expval {\nabla H}_t $ and $\expval {\nabla^2 H}_t $.
This can be achieved by either finding the differential equations they satisfy or determining their integrated form through a perturbative scheme.
In the first case, we derive the Heisenberg equation using the commutation relation~\cite{gadella1995moyal}
\begin{align} \label{m_105}
    & \comm{  q}{A  } = - i \hbar  J \nabla A , 
\end{align}
where $\nabla \mathcal A (z)$ is understood in a generalized function sense~\cite{de2017wigner}.
It can be shown that $A(t)$ evolves in the Heisenberg picture according to 
\begin{align} \label{m_118}
    & \dv{A_H (t)}{t}  = \Big( \pdv{A(t)}{t} \Big)_H + \expval{\nabla  H }_t ^\mathrm T J \nabla A_H \nonumber\\
    & - \tr{J \expval{\nabla ^2 H }_t \Big( (  q_H - \expval{   q }_t ) \nabla A_H + \frac{ i \hbar }{2}  J \nabla ^2 A_H \Big)  } .
\end{align}
The evolution of the expectation value is obtained from the Ehrenfest theorem,
\begin{align} \label{m_89}
    & \dv{ \expval{ A} _t }{t} = \expval{ \pdv{ A  }{t}}_t + \sigma (\expval{\nabla A}_t, \expval{\nabla H}_t)  \nonumber\\
    & - \tr{J \expval{ \nabla ^2 H }_t \Big ( \expval{( q - \expval{ q}_t) \nabla A }_t + \frac{i\hbar}{2}  J \expval{ \nabla ^2 A}_t \Big ) } ,
\end{align}
where $\sigma (a,\, b ) = b^\mathrm T J a $ is the standard symplectic form, with $\sigma (a,a) = 0$.
Replacing $ A $ by $ \nabla H$ or $ \nabla^2 H$ completes the SCQA equations, \cref{m_80}. 
These will, in turn, depend on higher-order derivatives $\expval{\nabla^n H}_t$ $( n \geq 3)$, creating a hierarchy of differential equations for the nonlinear contributions. 

\Cref{m_89} is well-defined when $\mathcal A (z)$ is a generalized function, provided that $\mathcal W_t (z)$ is continuously differentiable.
This is the case for Gaussian states
\begin{align} \label{m_64}
    & \mathcal  W_t ( z) = \det {2 \pi M_t }^{-1/2}  \nonumber\\
    & \exp{ -\frac{1}{2} ( z - \expval{ q}_t )^\mathrm T M_t^{-1} ( z -\expval{ q}_t ) } ,
\end{align}
where $M_t = \Lambda _t ^{-1} M_0 (\Lambda _t ^{-1} )^\mathrm T$.
Using \cref{m_64}, we can compute the term $\expval{ ( q - \expval{ q}_t) \nabla A}_t$ appearing in \cref{m_89} by applying the identity 
\begin{align} \label{m_85}
    & \expval{( q - \expval{ q}_t) A }_t = \Big( M_t + \frac{i\hbar}{2}J^\mathrm T \Big) \expval{ \nabla A}_t ,
\end{align}
which relates any operator to its gradient.
The quantity ${M_t + \frac{i\hbar}{2}J^\mathrm T} $ is associated with the uncertainty relations~\cite{de2017wigner}.
From the above result, we reduce \cref{m_89} to
\begin{align} \label{m_133}
    & \dv{ \expval{ A} _t }{t} = \expval{ \pdv{ A  }{t}}_t + \sigma (\expval{\nabla A}_t, \expval{\nabla H}_t)  \nonumber\\
    & - \tr{J \expval{ \nabla ^2 H }_t M_t \expval{ \nabla ^2 A}_t  } .
\end{align}
We are now set to prove that, for Gaussian states, $H$ is a constant of motion if it is time-independent. 
This is seen by taking ${A=H}$ in \cref{m_133}, from which $\sigma (\expval{\nabla H}_t, \expval{\nabla H}_t) = 0$ and 
\begin{align} \label{m_127}
    & \tr{J \expval{ \nabla ^2 H }_t M_t \expval{ \nabla ^2 H}_t } = 0 ,
\end{align}
due to the symmetry of the term multiplying $J$.
Then, 
\begin{align} \label{m_128}
    & \dv{ \expval{ H} _t }{t} = \expval{ \pdv{ H  }{t}}_t .
\end{align}
Stationary states can be found from the condition ${\dv{ \expval{ A} _t }{t} = 0}$, $\forall A $.
For Gaussian states, specifying the parameters $\expval q$ and $M$ is sufficient.
Substituting ${A = \diag[q^2] }$ in \cref{m_133} gives
\begin{align} \label{m_147}
    & \expval{\nabla^\mathrm T H} J \expval q - \tr {J \expval {\nabla^2 H} M} = 0,
\end{align}
which is verified by $\expval q = \expval {\nabla H}$ and $M = \expval {\nabla^2 H}$.

For $ H $ hermitian, it follows that $ H_{\text{SC}}$ is also hermitian, ensuring unitary SCQA dynamics and consequent entropy conservation.
Trace preservation results from $\Lambda_t $ being symplectic, a consequence of the symmetry property ${ B_t = B_t^\mathrm T}$~\cite{gadella1995moyal}.
For initial Gaussian states, the existence of $n$ dynamical invariants $\{ {\mathcal D _m,\, m=0,2,\dots, 2(n-1)} \}$ in the quadratic evolution results from~\cite{dodonov2000universal,dodonov2003theory}
\begin{align} \label{m_140}
    & \mathcal D(\mu; t) = \det{ M_t - \mu J } = \sum_{m=0}^{2N} \mathcal D_m \mu^m  = \mathcal D(\mu; 0)  \\
    & \mathcal D(\mu; 0) = \mathcal D(- \mu; 0) .
\end{align}
The coefficients $\mathcal D_m$ are independent of $B _t$ and $C_t$, which contain the self-consistent contributions to the Hamiltonian.
Some known conserved quantities are
\begin{align} \label{m_132}
    & \mathcal D_0 = \det M_t \\
    & \mathcal L_m = \tr { \Big ( M_t J^\mathrm T \Big )^m }, 
\end{align}
with $\mathcal L_{2k+1} = 0$ due to the symmetry properties of $M_t$ and $J$.
From these considerations, and taking into account $H$ conserved, we see the SCQA dynamics contains ${n-1}$ dynamical invariants.

We can also complete the SCQA equations by looking at the integrated form of $\expval{ \nabla H }_t$ and $\expval{ \nabla ^2 H }_t$.
An operator $A$ with continuously differentiable Weyl symbol $\mathcal A(z)$ can be expanded in a power series on the canonical vector $q$ and around some reference point ${\mathcal q =  q - \delta  q }$, ${ \mathcal q \in \mathbb R^{2n}}$, using the Taylor theorem~\cite{osborn1995moyal}
\begin{align} \label{m_09}
    &  A = e^{ ( q - \mathcal q) ^\mathrm T \nabla } \mathcal A (\mathcal q ) 
    = \sum_{ \alpha = 0}^\infty \frac{1}{\alpha !} \nabla ^\alpha \mathcal A (\mathcal q) (\delta q ) ^\alpha  .
\end{align}
The expression $\nabla ^\alpha \mathcal A (\mathcal q)$ indicates that derivatives first act on $\mathcal A(z)$, followed by evaluation at $\mathcal q$. 
The abbreviated multi-index notation is used here, with ${\alpha = (\alpha_1,\, \dots,\, \alpha _{2n})}$, ${\alpha ! = \alpha_1 ! \cdots \alpha _{2n}! }$, ${\nabla^\alpha = \nabla _1 ^{\alpha_1} \cdots \nabla _{2n} ^{\alpha_{2N}}}$ and ${\nabla _i ^{\alpha_i} := \pdv{^{\alpha_i}}{z_i ^{\alpha_i}}}$~\cite{folland2005higher}.
From \cref{m_09}, and using the commutation identity $\comm{\nabla_i}{\nabla_j}=0$, we can see that the Weyl symbol of $ \nabla^\alpha A $ is indeed $ \nabla^\alpha \mathcal A (z ) $. Also, $\expval A_t $ can be expanded as
\begin{align} \label{m_99}
    & \expval{ A}_t = e^{ - \nabla ^\mathrm T \mathcal q } \chi_t ( \nabla )  \mathcal A (\mathcal q) ,
\end{align}
where ${ \chi_t (a) \equiv \expval{ \exp{a ^\mathrm T  q}}_t }$ is the characteristic function of the state $ \rho_t $.
\Cref{m_99} is simplified for Gaussian states, whose characteristic function reads
\begin{align} \label{m_65}
    & \chi_t (a ) = \exp{ \frac{1}{2} a ^\mathrm T M_t a + a ^\mathrm T \expval{ q}_t } .
\end{align}
Choosing $\mathcal q = \expval q_t$ gives
\begin{align} \label{m_91}
    & \expval{ A}_t = e^{ \frac{1}{2} \nabla ^\mathrm T M_t \nabla } \mathcal A ( \expval q_t ) .
\end{align}
This result coincides with the Wick theorem in field theory~\cite{zee2010quantum}, where the Weyl symbol now represents the averaged function.
Substituting ${A=\nabla H}$ or ${ A=\nabla^2 H}$ and expanding the exponential in powers of $\nabla$ systematically includes higher-order contributions $\nabla^\alpha \mathcal H (\expval q_t )$ into the SCQA equations.
The truncation of the series to the zeroth order yields the Gaussian wave-packet dynamics~\cite{begusic2021finite,heller1975time}. 
It's worth noting that by reinterpreting the series terms in \cref{m_91} as Feynman diagrams, we could pursue partial summation over an infinite class of terms~\cite{stefanucci2013nonequilibrium}.
However, this approach lies beyond the scope of this work.

As an application of \cref{m_91}, observe that it solves \cref{m_147} for the stationary Gaussian state by 
\begin{align} \label{m_150}
    & e^{ \frac{1}{2} \nabla ^\mathrm T M \nabla } \Big( \sigma (\expval q, \nabla ) - \tr{M J \nabla ^2 } \Big) \mathcal H (\expval q) = 0 .
\end{align}
For the choice $\expval q = \expval {\nabla H}$ and $M = \expval {\nabla^2 H}$, it reduces to
\begin{align} \label{m_151}
    & \expval q = e^{ \frac{1}{2} \nabla ^\mathrm T M \nabla } \nabla \mathcal H (\expval q) \\
    & M = e^{ \frac{1}{2} \nabla ^\mathrm T M \nabla } \nabla^2 \mathcal H (\expval q) .
\end{align}
%

To study the system's response to a classical field $E(t)$, let's assume 
\begin{align} \label{m_74}
    & H (t) = H_c - E (t) V \hbar ,
\end{align}
where $H_c$ is independent of $\hbar$ and $t$.
The $N$th-order polarization reads \cite{mukamel1995principles}
\begin{align} \label{m_75}
    &  P ( \tau_1 ) = i^N \int _{t_0}^{\tau_1} \dd \tau_2\, \dots  \int _{t_0}^{\tau_N} \dd \tau_{N+1} \nonumber\\
    & E (\tau_2) \dots E(\tau_{N+1}) \, S (\tau_1,\, \dots,\, \tau_{N+1} )  ,
\end{align}
with ${ \tau_1 \geq \tau_2 \dots \geq \tau_{N+1} \geq t_0 }$ and the nonlinear response function defined by
\begin{align} \label{m_31}
    &  S (\tau_1 ,\, \dots,\, \tau_{N+1})  \nonumber\\
    & = \tr{ \rho_0 \comm{  \comm{ \dots \comm{ V (\tau_1 )}{ V (\tau_2)}} \dots }{ V (\tau_{N+1})} }.
\end{align}
The operator ${\rho_0 \equiv \rho (t_0)} $ corresponds to the equilibrium state  and remains invariant under the action of the evolution operator.
During the \textit{waiting time} $\tau_j$, the evolution is given by
\begin{align} \label{m_32}
    & V (\tau_j) =   U^\dagger (\tau_j)  V U (\tau_j), \quad U (\tau_j) = e^{-i/\hbar H_c \tau_j } .
\end{align}
Here, the SCQA is performed separately for each waiting time $\tau_j$, with $U (\tau_j) \approx U_{SC} (\tau_j)$.
Expanding the nested commutators in \cref{m_31} results~\cite{zagier1970expansion}
\begin{align} \label{m_76}
    &  S (\tau_1 ,\, \dots,\, \tau_{N+1}) \nonumber\\
    & = \sum_{k=1}^{N+1} (-1)^{k-1} \sum_{\sigma \in S_{N+1,k}} R ( \tau _{\sigma(1)},\, \dots,\, \tau _{\sigma(N+1)}) ,
\end{align}
where $S_{N+1,k}$ is the set of permutations $\sigma$ with ${ \sigma (1) > \dots > \sigma (k) = 1 < \sigma (k+1) < \dots < \sigma (N+1) }$, and 
\begin{align} \label{m_77}
    &  R ( \tau _{\sigma(1)},\, \dots,\, \tau _{\sigma(N+1)})  \nonumber\\
    & =  \tr { \rho_0 \prod _{j=1}^{N+1} U_{SC} ^\dagger (\tau _{\sigma(j)}) V  U_{SC} (\tau _{\sigma(j)}) } 
\end{align}
is the response function of a given permutation $\sigma$.
We now focus on this quantity and drop the permutation indices for simplicity.
It represents a generic real-time n-point correlation function~\cite{stefanucci2013nonequilibrium}.

From the Taylor series, \cref{m_09}, we can write the response function as
\begin{align} \label{m_78}
    &  R_N = \Xi _{N+1,0} \, \mathcal V (\mathcal q_1,\, \dots,\, \mathcal q_{N+1}) \\
    \label{m_112}
    & \Xi _{N+1,0} \equiv \expval{\prod_{j=1}^{N+1}  U_{j,\, SC} ^\dagger\, e^{\nabla_j ^\mathrm T (  q - \mathcal q_j) }\,  U_{j,\, SC} }_0 \\
    \label{m_113}
    &  \mathcal V (\mathcal q_1,\, \dots,\, \mathcal q_{N+1}) \equiv  \prod_{j=1}^{N+1} \mathcal V ( \mathcal q_j) ,
\end{align}
where the indices $j$ distinguish the waiting times, and there are $N+1$ canonical vectors with components $ q_{j\xi}$.
The differential operator $\Xi _{N+1,0} $ is calculated by means of the Weyl's characteristic function~\cite{de2017wigner}
\begin{align} \label{m_148}
    & M ( a ) \equiv \exp {\frac{i}{\hbar}  q ^\mathrm T   a } \\
    \label{m_149}
    & M ( a )   M ( b ) = e^{ \frac{i}{2\hbar} \sigma ( a,\,  b) }   M ( a +  b ) .
\end{align}
It can be shown that
\begin{align} \label{m_79}
    & \Xi _{N+1,0} = \chi_0 \Big ( \sum_{j=1}^{N+1} (\Lambda_j ^{-1} ) ^\mathrm T \nabla_j \Big ) \nonumber\\
    & \times e^ { -\frac{i\hbar }{2} \sum_{k>j=1}^{N+1} \sigma ((\Lambda_k ^{-1} ) ^\mathrm T \nabla_k ,\, (\Lambda_j ^{-1} ) ^\mathrm T \nabla_j ) - \sum_{j=1}^{N+1} \nabla_j ^\mathrm T (\Lambda_j ^{-1}  \Delta_j + \mathcal q_j ) } ,
\end{align}
where the information about the initial state enters through the characteristic function $\chi_0$.
From now on we refer to $\Xi _{N+1,0}$ as the \textit{response~operator}.

Before determining its explicit form, we point out a remarkable connection with the quantum optics characteristic function ${\chi [\rho] (\lambda) =\tr{\rho D(\lambda)} }$, where ${ D(\lambda)= \exp( i q^\mathrm T J^\mathrm T \lambda )}$ is the displacement operator~\cite{ferraro2005gaussian}. It relates to the Weyl's characteristic function through ${ D(\lambda) = M ( \hbar J^\mathrm T \lambda) }$. 
For exponential interactions, ${ V_j = e^{ a_j ^\mathrm T   q } }$~\cite{khidekel1995high}, the gradient $\nabla_j$ is replaced by $a_j$ in the response operator, which together with \cref{m_149} gives
\begin{align} \label{m_141}
    &  R_N = e^{  \varphi } \chi[\rho_0] \Big ( i J^\mathrm T  \sum_{j=1}^{N+1} (\Lambda_j ^{-1} ) ^\mathrm T a_j \Big )  \nonumber\\
    & \varphi = -\frac{i\hbar }{2} \sum_{k>j=1}^{N+1} \sigma ((\Lambda_k ^{-1} ) ^\mathrm T a_k ,\, (a_j ^{-1} ) ^\mathrm T a_j ) - \sum_{j=1}^{N+1} a_j ^\mathrm T \Lambda_j ^{-1}  \Delta_j .
\end{align}
This shows the equivalence between the nonlinear response and the quantum optics characteristic function up to a phase factor.
Consequently, for an unknown initial state, we can use its nonlinear response to perform a tomographic reconstruction via the identity~\cite{ferraro2005gaussian}
\begin{align} \label{m_142}
    &  \rho =  \Big( \frac{\hbar}{2\pi}\Big)^n \int_{\mathbb R^{2n}} \dd \lambda\, \chi [\rho] (\lambda) D ^\dagger (\lambda) .
\end{align}
This will require making ${\lambda = i J ^\mathrm T \sum_{j=1}^{N+1} (\Lambda_j ^{-1} ) ^\mathrm T a_j}$, which suggests using the waiting times $\tau _j$ as tunable parameters for scanning ${ \lambda \in \mathbb R^{2n}}$.

Evaluating the response operator, \cref{m_79}, over initial Gaussian states gives the nonlinear response function 
{\small
\begin{align} \label{m_116}
    & R _N = \exp { \frac{1}{2} \sum_{k=1}^{N+1} \nabla_k ^\mathrm T \Sigma _{kk} \nabla_k + \sum_{k>j=1}^{N+1} \nabla_j ^\mathrm T \Sigma _{jk} \nabla_k } \prod_{j=1}^{N+1} \mathcal V (\expval q_j ),
\end{align}
}
where $ \Sigma_{jk} = \Lambda _j ^{-1} \Sigma_0 ( \Lambda _k ^{-1} )^ \mathrm T $ and $ \Sigma_0  = M_0 + \frac{i\hbar}{2} J ^\mathrm T $.

The classical limit $\hbar \to 0$ for the nonlinear response, \cref{m_78,m_116}, requires careful consideration as it is known that divergences may arise due to anharmonicity~\cite{kryvohuz2006classical,wu2001linear,kryvohuz2005quantum}.
However, if we restrict to semiclassically admissible functions
\begin{align} \label{m_122}
    & \mathcal A (z) = \mathcal A_c (z) + \sum _{k=1}^\infty \frac{\hbar^k}{k!} a_k (z), 
\end{align}
whose Weyl symbol is asymptotically regular at ${\hbar = 0}$, e.g., the leading term is regarded as the classical counterpart of $A$~\cite{osborn1995moyal}, we can write 
\begin{align} \label{m_123}
    & \chi_0 (z) = \chi_c (z) + \mathcal O (\hbar) \\
    \label{m_124}
    & \mathcal V (z) = \mathcal V_c (z) + \mathcal O (\hbar), 
\end{align}
for which the nonlinear response is well defined as ${\hbar \to 0}$.
An example of the first equation is the characteristic function of the thermal state~\cite{akhundova1982wigner}.
However, \cref{m_123,m_124} do not hold in general and divergences might arise, for instance, when dealing with nonclassical quantum states.

In conclusion, this work has presented a self-consistent quadratic theory for nonlinear contributions in quantum dynamics, applicable to a broad class of Hamiltonians comprising composite qudit-boson systems. The theory emphasizes the existence of many conserved quantities, including the system Hamiltonian. Arbitrary-order nonlinear response functions have been calculated, their classical limit studied, and connections with quantum tomography are highlighted. 

\vspace{0.3cm}
The author would like to thank Filippo~Troiani and Raffaello~Bianco for sharing interesting and informative conversations.

The author declare that there are no conflicts of interest to disclose.

\bibliography{bib}

\clearpage
\onecolumngrid

\end{document}


\title{ Supplemental Material to Quantum dynamics in the self-consistent quadratic approximation}

\author{Frank Ernesto Quintela Rodriguez\,\orcidlink{0000-0002-9475-2267}}
\email[Email: ]{frank.quintelarodriguez@sns.it}
\affiliation{Scuola Normale Superiore, Piazza dei Cavalieri, 7, Pisa, 56126, Italy}

\begin{abstract}
\end{abstract}

\pacs{
}

\date{\today}

\maketitle

\tableofcontents
\makeatletter
\let\toc@pre\relax
\let\toc@post\relax
\makeatother

\section{Wigner-Weyl isomorphism}

The Wigner-Weyl isomorphism is a mapping that associates an operator $\hat f$ acting on the Hilbert space $L^2 (\mathbb R^n )$ to a distribution (generalized function) $\mathcal f(z) \equiv (\hat f)_W (z) $ defined in the phase space $z =(p,x) \in \mathbb R^{2n}$, through the transformation~\cite{gadella1993quantum}
\begin{align} \label{a_92}
    & \hat f =  \frac{1}{\sqrt \nu} \int_{\mathbb R^{2n}} \dd  \eta\, \dd  \sigma\, \mathscr  F ( \mathcal f ) (\eta,\, \sigma ) e^{i (\eta \hat p + \sigma \hat x)} ,
\end{align}
with $ \nu = (2 \pi \hbar )^{2n}$. 
This transformation is invertible and linear with respect to the sum of functions and products by complex numbers.
For the mapping being a homomorphism between the Hilbert and phase spaces, the Weyl symbol of the operator product ${ (\hat f \hat g )_W (z) = (f \star _\hbar g) (z) } $ is given through the Moyal product
\begin{align} \label{a_93}
    & (f \star _\hbar g ) (z) = \frac{1}{\sqrt \nu} \int_{\mathbb R^{4n}} \dd  u\, \dd  v\, \mathcal f(u) \mathcal g (v)  e^{i ( z J u + u J v + v J z )  } .
\end{align}
For analytic functions, the product can be computed using the Groenewold formula~\cite{de2017wigner}
\begin{align} \label{a_94}
    &  (\hat f \hat g )_W (z) = \mathcal f (z) e^{ - \frac{i\hbar }{2} \sigma (\overrightarrow \nabla, \overleftarrow \nabla ) } \mathcal g (z),
\end{align}
where ${ \sigma (a,\, b ) = b^\mathrm T J a }$ and the exponential (the ``Janus operator'') is understood as a power series, the arrows indicating the direction in which the derivative acts, i.e., $\overrightarrow \nabla$ will act on $\mathcal g (z)$.
It can be shown that
\begin{align} \label{a_95}
    & \tr{ \hat f \hat g} = \int_{\mathbb R^{2n}} \dd  z\, \mathcal f (z) \mathcal g (z)  = \int_{\mathbb R^{2n}} \dd  z\, \mathcal g (z) \mathcal f (z) = \tr{ \hat g \hat f},
\end{align}
which is cyclic as required. 
The Wigner-Weyl isomorphism can be defined by using the Grossmann-Royer operator
\begin{align} \label{a_96}
    & \hat \Pi ( z) \Psi (\zeta ) = 2^n e^{i  p ^\mathrm T (\zeta -  q )} \Psi (2  q - \zeta) ,
\end{align}
which verifies the identities
\begin{align} \label{a_97}
    & \tr \hat \Pi ( z) = 1 \\
    & \tr { \hat \Pi ( z) \hat \Pi ( z')} = 2^n \sqrt \nu \delta ( z -  z'),
\end{align}
where 
\begin{align} \label{a_91}
    & \tr \hat A = \int _{\mathbb R^n } \dd  x\, \matrixel{ x}{\hat A}{ x} .
\end{align}
Then, an operator $\hat f$ and its associated Weyl symbol $ \mathcal f ( z) $  are related through
\begin{subequations} \label{a_98}
\begin{align}
    &  \mathcal f ( z) = \tr{ \hat \Pi ( z) \hat f } \\
    & \hat f =  \frac{1}{\sqrt \nu} \int_{\mathbb R^{2n}} \dd  z\, \hat \Pi ( z)  \mathcal f ( z) .
\end{align}
\end{subequations}

\section{Gradient identity for Gaussian states } \label{a_34}

For initial Gaussian states, any operator relates to its gradient by
\begin{align} \label{a_77}
    & \expval{( q - \expval{ q}_t) A }_t = \Big( M_t + \frac{i\hbar}{2}J^\mathrm T \Big) \expval{ \nabla A}_t ,
\end{align}
where $M_t = \Lambda _t ^{-1} M_0 (\Lambda _t ^{-1} )^\mathrm T$.
We present two proofs, one using integration by parts and the other using the Taylor series expansion of an operator.

\begin{proof}
    \begin{align} \label{a_118}
        & \expval{ \delta q A }_t = \int_{\mathbb R^{2n}} \dd z\, (\delta q A)_W (z) \mathcal W_t (z)
        = \int_{\mathbb R^{2n}} \dd z\, \delta z \mathcal A (z) \mathcal W_t (z) + \frac{i\hbar}{2} J^\mathrm T \int_{\mathbb R^{2n}} \dd z\, \nabla \mathcal A (z) \mathcal W_t (z) 
    \end{align}
    from the Bopp shifts identity $ (\delta q A)_W (z) = \delta z \mathcal A (z) + \frac{i\hbar}{2} J^\mathrm T \nabla \mathcal A (z)$, which is valid in a generalized function sense~\cite{de2017wigner}.
For a Gaussian state
\begin{align} \label{a_119}
    & \mathcal  W_t ( z) = \det {2 \pi M_t }^{-1/2} \exp{ -\frac{1}{2} ( z - \expval{ q}_t )^\mathrm T M_t^{-1} ( z -\expval{ q}_t ) } ,
\end{align}
with $M_t = \Lambda _t ^{-1} M_0 (\Lambda _t ^{-1} )^\mathrm T$. Taking the gradient
\begin{align} \label{a_120}
    & \nabla \mathcal  W_t ( z) = - M_t ^{-1} \delta z \mathcal W (z),
\end{align}
where ${\delta z = z - \expval q_t}$ and we have used $M_t ^{-1} =( M_t ^{-1}) ^\mathrm T$.
Substituting \cref{a_120} into \cref{a_118} and integrating by parts gives 
\begin{align} \label{a_121}
    & \expval{ \delta q A }_t 
        = M_t \int_{\mathbb R^{2n}} \dd z\,  M_t ^{-1} \delta z \mathcal W_t (z) \mathcal A (z)  + \frac{i\hbar}{2} J^\mathrm T \int_{\mathbb R^{2n}} \dd z\, \nabla \mathcal A (z) \mathcal W_t (z) \nonumber\\
        & = - M_t \int_{\mathbb R^{2n}} \dd z\,  \nabla \mathcal W_t (z) \mathcal A (z)  + \frac{i\hbar}{2} J^\mathrm T \expval {\nabla A}_t \nonumber\\
        & = - M_t \Big( \mathcal W_t (z) \mathcal A (z) | _{-\infty}^\infty - \int_{\mathbb R^{2n}} \dd z\, \mathcal W_t (z) \nabla \mathcal A (z) \Big)   + \frac{i\hbar}{2} J^\mathrm T \expval {\nabla A}_t .
\end{align}
Since $\mathcal W_t (z=\pm \infty) =0 $, 
\begin{align} \label{a_122}
    & \expval{ \delta q A }_t  = M_t \int_{\mathbb R^{2n}} \dd z\, \mathcal W_t (z) \nabla \mathcal A (z)  + \frac{i\hbar}{2} J^\mathrm T \expval {\nabla A}_t = M_t \expval {\nabla A}_t+ \frac{i\hbar}{2} J^\mathrm T \expval {\nabla A}_t = \Big( M_t + \frac{i\hbar}{2}J^\mathrm T \Big) \expval {\nabla A}_t .
\end{align}
\end{proof}

\begin{proof}

\begin{enumerate}[A.]
    \item From the Groenewold formula, \cref{a_94}, we have 
        \begin{align} \label{a_107}
            &  (\delta q A )_W (z) = \delta z \mathcal A (z) + \frac{i \hbar }{2} J^\mathrm T \nabla \mathcal A (z) .
        \end{align}
    \item For initial Gaussian states the expectation value is given by 
        \begin{align} \label{a_130}
            & \expval{ A}_t = e^{ \frac{1}{2} \nabla ^\mathrm T M_t \nabla } \mathcal A ( \expval q_t ) ,
        \end{align}
        then substituting \cref{a_107} results
    \begin{align} \label{a_108}
        & \expval{ \delta q A}_t = e^{ \frac{1}{2} \nabla ^\mathrm T M_t \nabla } \Big( \delta z \mathcal A (z) + \frac{i \hbar }{2} J^\mathrm T \nabla \mathcal A (z) \Big) \Big |_{z= \expval q_t } 
        = e^{ \frac{1}{2} \nabla ^\mathrm T M_t \nabla } \big( \delta z \mathcal A (z) \big) \Big |_{z= \expval q_t } + \frac{i \hbar }{2} J^\mathrm T \expval{ \nabla A} .
    \end{align}
\item To evaluate the first term in the \textit{r.h.s} of \cref{a_108}, we denote $ \innerp*{\nabla, M \nabla}:= \nabla ^\mathrm T M_t \nabla $ and prove the identity
    \begin{align} \label{a_109}
        & \frac{1}{2} \comm{\innerp{\nabla, M \nabla}^k }{z} = k \innerp{\nabla, M \nabla} ^{k-1} M \nabla ,\quad k \geq 1.
    \end{align}
        \begin{enumerate}
            \item[$k=1$] 
                \begin{align} \label{a_110}
                    & \comm{\nabla_\mu \nabla_\nu }{z_\alpha }f = \nabla _{\mu \nu } ( z_\alpha f ) - z_\alpha \nabla _{\mu \nu } f = 
                    \nabla _{\mu} (\delta_{\alpha \nu } f + z_\alpha \nabla _{\nu} f )  - z_\alpha \nabla _{\mu \nu } f 
                    =  \delta_{\alpha \nu } \nabla _{\mu} f + \delta_{\alpha \mu } \nabla _{\nu} f .
                \end{align}
                Then we have $\comm{\nabla_\mu \nabla_\nu }{z_\alpha } = \delta_{\alpha \nu } \nabla _{\mu} + \delta_{\alpha \mu } \nabla _{\nu} $.
                From this results
                \begin{align} \label{a_111}
                    & \frac{1}{2} \sum_{\mu \nu} M_{\mu \nu } \comm{\nabla_\mu \nabla_\nu }{z_\alpha } = 
                    \frac{1}{2} \Big ( \sum_{\mu } M_{\mu \alpha } \nabla _\mu  + \sum_{\nu } M_{\alpha \nu } \nabla _\nu \Big ) = \Big( \frac{1}{2} (M^\mathrm T + M) \nabla \Big) _\alpha .
                \end{align}
                Assuming $M^\mathrm T = M$ we get
                \begin{align} \label{a_112}
                    & \frac{1}{2} \comm{\innerp{\nabla, M \nabla }}{z} = M \nabla ,
                \end{align}
                which is \cref{a_109} for $k=1$.

            \item[$k\Rightarrow k + 1$] 

                \begin{align} \label{a_113}
                    & \frac{1}{2} \comm{\innerp{\nabla, M \nabla }^{k+1}}{z} 
                    = \frac{1}{2} \innerp{\nabla, M \nabla }^{k+1}z - \frac{1}{2} z \innerp{\nabla, M \nabla }^{k+1} \nonumber\\
                    & =  \innerp{\nabla, M \nabla } \Big( \frac{1}{2} z \innerp{\nabla, M \nabla }^k  + k \innerp{ \nabla, M \nabla }^{k-1} M \nabla \Big)  - \frac{1}{2} z \innerp{\nabla, M \nabla }^{k+1} \nonumber\\
                    & = \Big( \frac{1}{2} z \innerp{\nabla, M \nabla } + M \nabla \Big) \innerp{\nabla, M \nabla }^k  + k \innerp{ \nabla, M \nabla }^k M \nabla  - \frac{1}{2} z \innerp{\nabla, M \nabla }^{k+1} \nonumber\\
                    & = (k + 1 ) \innerp{ \nabla, M \nabla }^k M \nabla .
                \end{align}
                Together \cref{a_112,a_113} prove \cref{a_109}.
        \end{enumerate}
        
    \item Expanding the exponential in power series,
        \begin{align} \label{a_114}
            & e^{ \frac{1}{2} \nabla ^\mathrm T M_t \nabla } \big( \delta z \mathcal A (z) \big) \Big |_{z= \expval q_t }
            = \sum_{k=0}^\infty \frac{1}{k! 2^{k-1}} \frac{1}{2} \innerp{\nabla, M \nabla }^k \big( \delta z \mathcal A (z) \big) \Big |_{z= \expval q_t },
        \end{align}
        and from the commutation relation in \cref{a_109} we get
        \begin{align} \label{a_115}
            & =  \big( \delta z \mathcal A (z) \big) \Big |_{z= \expval q_t } 
            + \sum_{k=1}^\infty \frac{1}{k! 2^k } \big( \delta z \innerp{\nabla, M \nabla }^k  \mathcal A (z) \big) \Big |_{z= \expval q_t } 
            + \sum_{k=1}^\infty \frac{1}{k!} \frac{1}{2^{k-1}} k \innerp{\nabla, M \nabla }^{k-1} M \nabla \mathcal A (z)  \Big |_{z= \expval q_t } .
        \end{align}
        Evaluating $z= \expval q_t \Rightarrow \delta z = 0$, so \cref{a_114,a_115} reduces to
        \begin{align} \label{a_116}
            & e^{ \frac{1}{2} \nabla ^\mathrm T M_t \nabla } \big( \delta z \mathcal A (z) \big) \Big |_{z= \expval q_t }
            = M \sum_{k=0}^\infty \frac{1}{k!} \frac{1}{2^k} \innerp{\nabla, M \nabla }^k \nabla \mathcal A (z) \Big |_{z= \expval q_t }
            =  M e^{ \frac{1}{2} \nabla ^\mathrm T M_t \nabla } \nabla A (\expval q_t ) = M \expval {\nabla A}_t .
        \end{align}
    From \cref{a_108,a_116} we finally get 
    \begin{align} \label{a_117}
        & \expval{ \delta q A}_t = M \expval {\nabla A}_t + \frac{i \hbar }{2} J^\mathrm T \expval{ \nabla A}_t,
    \end{align}
    which is nothing but the desired identity, \cref{a_77}. 

\end{enumerate}
\end{proof}

\section{The basic commutation relation}

Let's prove the commutation identity (also Eq.~(2.13) in \cite{gadella1995moyal})
\begin{align} \label{a_27}
    & \comm{  q}{A  } = - i \hbar  J \nabla A 
\end{align}
using both the Groenewold formula, \cref{a_94}, and the Taylor series expansion of operators.

\begin{proof}

    \begin{align} \label{a_133}
        & (q_\mu A )_W = z_\mu e^{ - \frac{i\hbar }{2} \sigma (\overrightarrow \nabla, \overleftarrow \nabla ) } \mathcal A (z) 
        = z_\mu \mathcal A (z) - \frac{i\hbar }{2} z_\mu \overleftarrow \nabla ^\mathrm T J \overrightarrow \nabla  \mathcal A (z)  
        = z_\mu \mathcal A (z) - \frac{i\hbar }{2} \delta_{\mu k} J_{k \nu } \nabla_\nu \mathcal A (z)  \\
        & (A q_\mu )_W = z_\mu \mathcal A (z) - \frac{i\hbar }{2} \mathcal A (z) \overleftarrow \nabla ^\mathrm T J \overrightarrow \nabla z_\mu
        = z_\mu \mathcal A (z) - \frac{i\hbar }{2} \nabla_\nu \mathcal A (z) J_{\nu k }  \delta_{\mu k} 
        = z_\mu \mathcal A (z) + \frac{i\hbar }{2} \delta_{\mu k} J_{k \nu } \nabla_\nu \mathcal A (z)   ,
    \end{align}
    where we have used $J_{\nu k } = - J_{k \nu }$. Then we have 
    \begin{align} \label{a_134}
        & (q_\mu A )_W - (A q_\mu )_W = - i\hbar J_{\mu \nu } \nabla_\nu \mathcal A (z) .
    \end{align}
    Back to the Hilbert space, \cref{a_134} coincides with \cref{a_27}.

\end{proof}

\begin{proof}

The canonical commutation relation for the operators ordering $  q = ( p_1,\, \dots,\,  p_n,\,  x_1,\, \dots,\,  x_n) ^\mathrm T $ is~\cite{dodonov2003theory}
\begin{align} \label{a_29}
    & \comm{  q_\xi}{ q_\eta} = - i \hbar  J _{\xi \eta} .
\end{align}
where $  J = \begin{pmatrix}  0 &  I \\ -  I &  0 \end{pmatrix}$ is the standard symplectic matrix, $ I \in \mathcal M (\mathbb R^n) $, and $ J^\mathrm T = J^{-1} = - J $. 
For $ u = \nabla ^\mathrm T \delta  q $,
\begin{align} \label{a_28}
    & \comm{ u}{ q_\eta } = \nabla_\xi \comm{\delta  q_\xi}{ q_\eta} = - i\hbar \nabla_\xi  J_{\xi \eta} \Rightarrow 
    \comm{q}{u} = i \hbar \nabla  J .
\end{align}
Now we prove that
\begin{align} \label{a_71}
    & \comm{ q}{ u^n } = i \hbar \nabla  J n  u ^{n-1} = - i \hbar  J  \nabla n  u ^{n-1} ,\quad n\geq 1.
\end{align}
\begin{proof}
    Let's proceed by induction. The case $n=1$ is given by \cref{a_28}.
    The inductive step $n-1 \Rightarrow n $ reads:
    \begin{align} \label{a_72}
        &  \comm{q}{u^n} = q u^{n-1} u - u^n q = (u^{n-1} q + i \hbar \nabla J (n-1) u^{n-2}) u - u^n q \nonumber\\
        & = u^{n-1} (u q + i \hbar \nabla J) + i \hbar \nabla J (n-1) u^{n-1} - u^n q = i \hbar \nabla J n u^{n-1}
    \end{align}

\end{proof}
Using \cref{a_71}, the commutator with the exponential reads
\begin{align} \label{a_31}
    & \comm{  q}{ e^{ u} } = \sum_{n=0}^\infty \frac{1}{n!} \comm{  q}{ u ^n} = - i \hbar  J \nabla \sum_{n=1}^\infty \frac{ n  u ^{n-1} }{n!} = - i \hbar  J \nabla e^{ u}
\end{align}
Comparing with the Taylor theorem gives $\comm{  q}{ A  } = - i \hbar  J \nabla A  $.
\end{proof}

\section{Heisenberg equation for the SCQA}

We start with the commutation identity \cref{a_27},
\begin{align} \label{a_132}
    & \comm{  q}{A  } = - i \hbar  J \nabla A .
\end{align}
Then 
\begin{align} \label{a_19}
    & \sum_{\eta=1}^{2n}  C_\eta \comm{   q_\eta }{A}  = i \hbar \nabla ^\mathrm T A   J  C \\
\label{a_41}
    & \frac{1}{2} \sum_{\eta,\mu =1}^{2n}  B_{\eta \mu} \comm{   q_\eta   q_\mu }{A} = - \frac{i \hbar}{2} (    q^\mathrm T  B  J \nabla A  - \nabla ^\mathrm T A   J  B   q) 
\end{align}

\begin{proof}

\begin{align} \label{a_40}
    & \comm{   q_\eta   q_\mu }{A} =    q_\eta   q_\mu A - A   q_\eta   q_\mu 
    =  q_\eta ( A   q_\mu - i\hbar  J _{\mu\xi}\nabla _\xi A ) - A   q_\eta   q_\mu  \nonumber\\
    & = ( A   q_\eta - i\hbar  J _{\eta\xi} \nabla _\xi A)   q_\mu - i\hbar  q_\eta   J _{\mu \xi}\nabla _\xi A - A   q_\eta   q_\mu \nonumber\\
    & =  - i\hbar  q_\eta   J _{\mu\xi}\nabla _\xi A - i\hbar  J _{\eta\xi}\nabla _\xi A q_\mu
\end{align}
Reordering the above terms and adding the products with $ \frac{1}{2}  B_{\eta \mu}$ gives
\begin{align} \label{a_40}
    & \frac{1}{2} \sum_{\eta,\mu =1}^{2n}  B_{\eta \mu} \comm{   q_\eta   q_\mu }{A}
    = - \frac{1}{2} \sum_{\eta,\mu =1}^{2n}  B_{\eta \mu}( i\hbar  q_\eta   J _{\mu\xi} \nabla _\xi A + i\hbar  J _{\eta\xi}\nabla _\xi A   q_\mu ) = - \frac{i \hbar}{2} (  q^\mathrm T  B  J \nabla A  - \nabla ^\mathrm T A   J  B   q)
\end{align}
\end{proof}

Now, let's work on the first term in the \textit{rhs} of \cref{a_40},
\begin{align} \label{a_26}
    &  q_\eta  J_{\mu\xi} \nabla_\xi A =  J_{\mu \xi} \nabla_\xi A  q_\eta + i\hbar  J_{\eta \nu} \nabla _{\nu \xi} A   J_{\xi \mu} .
\end{align}
Applying $ \sum_{\eta,\mu =1}^{2n}  B_{\eta \mu} $ in both sides of the equality,
\begin{subequations} \label{a_63}
\begin{align} 
    & \sum_{\eta,\mu =1}^{2n}  B_{\eta \mu} q_\eta  J_{\mu\xi} \nabla_\xi A  = q^\mathrm T  B  J \nabla A  \\
    & \sum_{\eta,\mu =1}^{2n} B_{\eta \mu} (J_{\mu \xi} \nabla_\xi A  q_\eta + i\hbar  J_{\eta \nu} \nabla _{\nu \xi} A   J_{\xi \mu}) = - \nabla ^\mathrm T A   J  B^\mathrm T   q + i \hbar \tr{  B^\mathrm T  J \nabla ^2 A   J } .
\end{align}
\end{subequations}
From \cref{a_26,a_63} results
\begin{align} \label{a_64}
    & q^\mathrm T  B  J \nabla A = - \nabla ^\mathrm T A   J  B^\mathrm T   q + i \hbar \tr{  B^\mathrm T  J \nabla ^2 A   J } .
\end{align}
\Cref{a_64} and $ B^\mathrm T = B$ symmetric allows to rewrite \cref{a_41} as
\begin{align} \label{a_39}
    & \frac{1}{2} \sum_{\eta,\mu =1}^{2n}  B_{\eta \mu} \comm{   q_\eta   q_\mu }{A} = i\hbar \nabla ^\mathrm T A  J B q + \frac{\hbar^2}{2} \tr{  B J \nabla ^2 A   J } .
\end{align}
The commutator with $H _{\text{SC}} (t) = \frac{1}{2}  q ^\mathrm T  B_t  q +  q ^\mathrm T  C_t $ reads
\begin{align} \label{a_25}
    & \comm{A  }{ H _{\text{SC}} (t)} = \comm{A  }{\frac{1}{2}  q ^\mathrm T  B_t  q +  q ^\mathrm T  C_t  } = -\sum_{\eta=1}^{2n}  C_\eta \comm{   q_\eta }{A} - \frac{1}{2} \sum_{\eta,\mu =1}^{2n}  B_{\eta \mu} \comm{   q_\eta   q_\mu }{A} \nonumber\\
    & = - i \hbar \nabla ^\mathrm T A   J  C\nonumber - i\hbar \nabla ^\mathrm T A  J B q - \frac{\hbar^2}{2} \tr{  B J \nabla ^2 A   J } = - i \hbar \nabla ^\mathrm T A J (C +B q) - \frac{\hbar^2}{2} \tr{  B J \nabla ^2 A   J }.
\end{align}
From 
\begin{subequations} \label{a_66}
\begin{align} 
    &  B_t = \expval{ \nabla ^2 H }_t,\quad B_t = B_t ^\mathrm T \\
    &  C_t = \expval{ \nabla H}_t -   \expval{ \nabla ^2 H } _t  \expval{  q} _t 
\end{align}
\end{subequations}
we have 
\begin{align} \label{a_67}
    & C +B q = \expval{ \nabla H}_t +  \expval{ \nabla ^2 H } _t  ( q - \expval{  q} _t ),
\end{align}
so the commutator gives 
\begin{align} \label{a_65}
    & \comm{A  }{ H _{\text{SC}} (t)} = - i \hbar \nabla ^\mathrm T A   J \expval{\nabla  H }_t 
     - i \hbar \nabla ^\mathrm T A   J \expval{\nabla ^2 H }_t (  q - \expval{   q }_t ) 
     - \frac{\hbar^2 }{2} \tr{  J \expval{\nabla ^2 H }_t  J \nabla ^2 A } .
\end{align}

Now, 
\begin{align} \label{a_99}
    & \frac{1}{i\hbar} \comm{A}{H _{\text{SC}} (t)} = - \nabla ^\mathrm T A   J \expval{\nabla  H }_t 
     - \nabla ^\mathrm T A   J \expval{\nabla ^2 H }_t (  q - \expval{   q }_t ) 
     + \frac{ i \hbar }{2} \tr{  J \expval{\nabla ^2 H }_t  J \nabla ^2 A } .
\end{align}

Writing the commutation relation \cref{a_27} in the form $ q_k O_\mu = O_\mu q_k - i \hbar J_{k \sigma } \nabla _\sigma O_\mu $ gives 
\begin{align} \label{a_100}
    & \nabla _\mu A J _{\mu \sigma } B_{\sigma k} \delta q_k  
    = J _{\mu \sigma } B_{\sigma k} \nabla _\mu A \delta q_k
    = J _{\mu \sigma } B_{\sigma k} ( \delta q_k \nabla _\mu A  + i \hbar J_{k \sigma } \nabla _{\sigma \mu} A ) \nonumber\\
    & = - \delta q_k B_{k \sigma }  J _{ \sigma \mu }  \nabla _\mu A  + i \hbar J _{\mu \sigma } B_{\sigma k} J_{k \sigma } \nabla _{ \sigma \mu } A
    = - \delta q ^\mathrm T B J \nabla A + i \hbar \tr{ J B J \nabla^2 A} ,
\end{align}
where we have used $B = B^\mathrm T$ and $J^\mathrm T = - J$. 
Substituting \cref{a_100} in \cref{a_99} gives 
\begin{align} \label{a_138}
    & \frac{1}{i \hbar } \comm{A  }{ H _{\text{SC}} (t)} = - \nabla ^\mathrm T A   J \expval{\nabla  H }_t 
     + (  q - \expval{   q }_t ) ^\mathrm T \expval{\nabla ^2 H }_t J \nabla A
     - \frac{ i \hbar }{2} \tr{  J \expval{\nabla ^2 H }_t  J \nabla ^2 A } \nonumber\\
     & =  \expval{\nabla  H }_t ^\mathrm T J \nabla A - \tr{J \expval{\nabla ^2 H }_t (  q - \expval{   q }_t ) \nabla A}
     - \frac{ i \hbar }{2} \tr{  J \expval{\nabla ^2 H }_t  J \nabla ^2 A } \nonumber\\
     & =  \expval{\nabla  H }_t ^\mathrm T J \nabla A - \tr{J \expval{\nabla ^2 H }_t \Big( (  q - \expval{   q }_t ) \nabla A + \frac{ i \hbar }{2}  J \nabla ^2 A \Big)  } ,
\end{align}
where we have used ${a^\mathrm T M b = \tr{M^\mathrm T a b}}$.

In the Heisenberg picture, ${A_H}:= U_{SC}^\dagger (t)  A(t) U_{SC}(t)$, an operator evolves under the self-consistent Hamiltonian $H_{SC}(t)$ according to
\begin{align} \label{a_105}
    & \dv{A_H (t)}{t}  = \Big( \pdv{A(t)}{t} \Big)_H + \frac{1}{i \hbar } \comm{A  }{ H_{SC} (t) }_H.
\end{align}
Substituting \cref{a_138} into \cref{a_105} gives
\begin{align} \label{a_137}
    & \dv{A_H (t)}{t}  = \Big( \pdv{A(t)}{t} \Big)_H + \expval{\nabla  H }_t ^\mathrm T J \nabla A_H - \tr{J \expval{\nabla ^2 H }_t \Big( (  q_H - \expval{   q }_t ) \nabla A_H + \frac{ i \hbar }{2}  J \nabla ^2 A_H \Big)  } ,
\end{align}
which is the Heisenberg equation for the SCQA.

\section{Trace and entropy preservation}

\begin{enumerate}[A.]
    \item $\Lambda_t$ is symplectic independently of the specific form of $B_t$ and $C_t$ (see proof in~\cite{gadella1995moyal}), so $\det {\Lambda _t} =1 $.
    \item 
\begin{align} \label{a_106}
    & \tr \rho_t = \int \dd z\, \mathcal W _t (z) = \int \dd z\, \mathcal W (\Lambda_t z + \Delta_t ) = \det {\Lambda _t} ^{-1} \int \dd z\,  \mathcal W (z) = \tr \rho_0 = 1.
\end{align}
\item If $ H=H ^\dagger \Rightarrow \nabla ^\alpha H = (\nabla ^\alpha H )^\dagger \Rightarrow H_{SC} (t) = H_{SC} ^\dagger (t)  $, so the dynamics is unitary. Then the entropy verifies
    \begin{align} \label{m_139}
        & S_t = - \tr{\rho_t \log \rho_t } = - \tr{ U _t \rho_0  U^\dagger _t \log ( U _t \rho_0  U^\dagger _t ) } 
    = -\tr{\rho_0 \log \rho_0 } = S_0 ,
    \end{align}
    where we have used the cyclic property of the trace and $U_t ^\dagger U_t = I$.

\end{enumerate}
 
\section{Nonlinear response function} \label{a_12}

We aim to compute the nonlinear response
\begin{align} \label{a_13}
    & \mathcal R_N = \expval{\prod_{j=1}^{N+1} U_{j, SC} ^\dagger e^{\nabla_j ^\mathrm T ( q - \mathcal q_j) } U_{j,SC} }_0 \prod_{j=1}^{N+1} \mathcal V ( \mathcal q_j ) \nonumber\\
    & = \expval{ \prod_{j=1}^{N+1} e^{ \nabla_j ^\mathrm T \Lambda_j ^{-1}  q } }_0 
    e^{- \sum_{j=1}^{N+1} \nabla_j ^\mathrm T (\Lambda_j ^{-1}  \Delta_j + \mathcal q_j ) }
    \prod_{j=1}^{N+1} \mathcal V ( \mathcal q_j )  \equiv \Xi_{N+1,0} \, \mathcal V(\mathcal q_1, \dots, \mathcal q_{N+1}).
\end{align}
The fundamental quantity in the above expression is $\Xi_{N+1,0} $, which contains the dynamical information through the differential operators $\nabla _j$ and the initial conditions in the expectation value.
From now on we refer to $\Xi _{N+1,0}$ as the \textit{response~operator}.
The product of exponentials appearing in it can be calculated using the Weyl's characteristic function~\cite{de2017wigner}
\begin{align} \label{a_82}
    & M ( a ) \equiv \exp {\frac{i}{\hbar}  a ^\mathrm T   q } \\
\label{a_81}
    &   M ( a )   M ( b ) = e^{ \frac{i}{2\hbar} \sigma ( a,\,  b) }   M ( a +  b ).
\end{align}
It verifies the identity
\begin{align} \label{a_14}
    &  \prod_{j=1}^{N+1} M ( a_j) = e^{\frac{i}{2\hbar} \sum_{k>j=1}^{N+1} \sigma ( a_k,\,  a_j) } M \Big( \sum_{j=1}^{n+1}  a_j \Big ) .
\end{align}
\begin{proof}

The starting point is the recurrence relation
\begin{align} \label{a_78}
    & e^{\frac{i}{2\hbar} \varphi _{j+1} } M (b_{j+1}) =  M (a_{j+1}) e^{\frac{i}{2\hbar} \varphi _j } M (b_j) = 
    e^{\frac{i}{2\hbar} ( \sigma (a_{j+1},b_j) + \varphi _j) } M (a_{j+1} + b_j ),
\end{align}
from which result the linear recurrence relations
\begin{align} \label{a_79}
    & b_{j+1} = a_{j+1} + b_j, \quad b_0 = 0 \\
    & \varphi _{j+1} = \sigma (a_{j+1},b_j) + \varphi _j, \quad \varphi _0 = 0.
\end{align}
The solution is given by
\begin{align} \label{a_80}
    & b _n = \sum_{j=1}^n a _j \\
    & \varphi _n = \sum_{k=1}^n \sigma (a _k,b _{k-1}) = \sum_{k > j=1}^n \sigma (a _k, a _j) .
\end{align}
\end{proof}

Using the definition of Weyl's characteristic function, \cref{a_82}, we can rewrite the exponential of the gradient as
\begin{align} \label{a_15}
    & e^{ \nabla_j ^\mathrm T \Lambda_j ^{-1}  q} = M ( -i\hbar (\Lambda_j ^{-1} ) ^\mathrm T \nabla_j ) ,
\end{align}
from which follows
\begin{align} \label{a_16}
    &  \prod_{j=1}^{N+1} e^{ \nabla_j ^\mathrm T \Lambda_j ^{-1}  q } = \prod_{j=1}^{N+1} M ( -i\hbar (\Lambda_j ^{-1} ) ^\mathrm T \nabla_j ) 
    = e^{ -\frac{i\hbar }{2} \sum_{k>j=1}^{N+1} \sigma ((\Lambda_k ^{-1} ) ^\mathrm T \nabla_k ,\, (\Lambda_j ^{-1} ) ^\mathrm T \nabla_j ) }
    M \Big( -i\hbar \sum_{j=1}^{N+1} (\Lambda_j ^{-1} ) ^\mathrm T \nabla_j \Big ) .
\end{align}
Taking the expectation value 
\begin{align} \label{a_17}
    &  \expval{ \prod_{j=1}^{N+1} e^{ \nabla_j ^\mathrm T \Lambda_j ^{-1}  q } }_0 =
    \exp{ -\frac{i\hbar }{2} \sum_{k>j=1}^{N+1} \sigma ((\Lambda_k ^{-1} ) ^\mathrm T \nabla_k ,\, (\Lambda_j ^{-1} ) ^\mathrm T \nabla_j ) }
    \expval{ M \Big( -i\hbar \sum_{j=1}^{N+1} (\Lambda_j ^{-1} ) ^\mathrm T \nabla_j \Big ) }_0 \nonumber\\
    & = \exp { -\frac{i\hbar }{2} \sum_{k>j=1}^{N+1} \sigma ((\Lambda_k ^{-1} ) ^\mathrm T \nabla_k ,\, (\Lambda_j ^{-1} ) ^\mathrm T \nabla_j ) }
    \chi_0 \Big ( \sum_{j=1}^{N+1} (\Lambda_j ^{-1} ) ^\mathrm T \nabla_j \Big ) 
\end{align}
Including the product with the exponential $ e^{- \sum_{j=1}^{N+1} \nabla_j ^\mathrm T (\Lambda_j ^{-1}  \Delta_j + \mathcal q_j ) }$ appearing in \cref{a_13}, we can write the response~operator as
\begin{align} \label{a_83}
    & \Xi _{N+1,0} = \chi_0 \Big ( \sum_{j=1}^{N+1} (\Lambda_j ^{-1} ) ^\mathrm T \nabla_j \Big ) 
    \exp { -\frac{i\hbar }{2} \sum_{k>j=1}^{N+1} \sigma ((\Lambda_k ^{-1} ) ^\mathrm T \nabla_k ,\, (\Lambda_j ^{-1} ) ^\mathrm T \nabla_j ) - \sum_{j=1}^{N+1} \nabla_j ^\mathrm T (\Lambda_j ^{-1}  \Delta_j + \mathcal q_j ) } 
\end{align}
For initial Gaussian states, the characteristic function has the simple 
\begin{align} \label{a_131}
    & \chi_0 (a ) = \exp{ \frac{1}{2} a ^\mathrm T M_0 a + a ^\mathrm T \expval{ q}_0 } ,
\end{align}
then
\begin{align} \label{a_84}
    & \chi_G \Big ( \sum_{j=1}^{N+1} (\Lambda_j ^{-1} ) ^\mathrm T \nabla_j \Big ) = 
    \exp{\frac{1}{2} \sum_{k,j=1}^{N+1} \nabla_j ^\mathrm T M_{jk}  \nabla_k  
    + \Big( \sum_{j=1}^{N+1} \nabla_j ^\mathrm T \Lambda_j ^{-1} \Big) \expval q_0 } ,
\end{align}
where $ M_{jk} \equiv \Lambda_j ^{-1} M_0 (\Lambda_k ^{-1} ) ^\mathrm T $, and  $M_{jk} = M_{kj}^\mathrm T $ symmetric whenever $M_0 = M_0 ^\mathrm T $ is symmetric.
From the symmetry argument follows the identity
\begin{align} \label{m_114}
    & \frac{1}{2} \sum_{k,l =1} ^{N+1} a_k ^\mathrm T M _{kl} a_l = \sum_{k>l =1} ^{N+1} a_k ^\mathrm T M _{kl} a_l + \frac{1}{2} \sum_{k =1} ^{N+1} a_k ^\mathrm T M _{kk} a_k ,
\end{align}
which we use in conjunction with \cref{a_83,a_84} to obtain
\begin{align} \label{a_85}
    & \Xi _{N+1,G} = \exp { \sum_{k>j=1}^{N+1} \nabla_j ^\mathrm T \Sigma _{jk} \nabla_k + \frac{1}{2} \sum_{k=1}^{N+1} \nabla_k ^\mathrm T \Sigma _{kk} \nabla_k -  \sum_{j=1}^{N+1} \nabla_j ^\mathrm T \Lambda _j ^{-1} ( \Lambda _j \mathcal q_j + \Delta _j - \expval q _0 ) } ,
\end{align}
where $ \Sigma_{jk} = \Lambda _j ^{-1} \Sigma_0 ( \Lambda _k ^{-1} )^ \mathrm T $ and $ \Sigma_0  = M_0 + \frac{i\hbar}{2} J ^\mathrm T $.

\begin{proof}
    Use: i) $ \nabla ^T J \nabla = \sigma ( \nabla, \nabla) = \comm{\nabla _p }{\nabla _x } = 0$. 
    ii) $ \Lambda _k $ symplectic, so $ \Lambda _k ^{-1} J ( \Lambda _k ^{-1}) ^\mathrm T = J $.
\end{proof}

Choosing $\mathcal q_j = \expval q _j = \Lambda_j ^{-1} ( \expval q_0 - \Delta_j )$ the linear term in \cref{a_85} vanishes, resulting 
\begin{align} \label{a_86}
    & \Xi _{N+1,G} = \exp { \frac{1}{2} \sum_{k=1}^{N+1} \nabla_k ^\mathrm T \Sigma _{kk} \nabla_k + \sum_{k>j=1}^{N+1} \nabla_j ^\mathrm T \Sigma _{jk} \nabla_k } .
\end{align}

\bibliography{bib}